\documentclass[twocolumn,showpacs,preprintnumbers,amsmath,amssymb]{revtex4}

\usepackage{graphicx}
\usepackage{dcolumn}
\usepackage{bm}

\begin{document}

\preprint{APS}

\title{Verification of Maxwell-Boltzmann distribution with Big-Bang Nucleosyntheis theory}

\author{S.Q.~Hou$^{1,2,3}$}
\author{J.J.~He$^1$}
\email{jianjunhe@impcas.ac.cn}
\author{other collaborators$^6$}
\affiliation{$^1$Key Laboratory of High Precision Nuclear Spectroscopy and Center for Nuclear Matter Science, Institute of Modern Physics, Chinese Academy of Sciences, Lanzhou 730000, China}%
\affiliation{$^2$University of Chinese Academy of Sciences, Beijing 100049, China}%
\affiliation{$^3$China Institute of Atomic Energy, Beijing 102413, China}%
\affiliation{$^6$International collaboration}%

\date{\today}

\begin{abstract}
The current Big-Bang Nucleosynthesis (BBN) model has been constructed based on a nuclear reaction network operating with thermal reactivities of 
Maxwell-Boltzmann (MB) distribution plasma. However, does the classical MB distribution still hold for the extremely high-temperature (in order of 10$^9$ K) 
plasma involved in the Big-Bang environment? 
In this work, we have investigated the impact of non-extensive Tsallis statistics (in $q$-Guassian distribution) on the thermonuclear reaction rates.
We show for the first time that the reverse rates are extremely sensitive to the non-extensive $q$ parameter. Such sensitivity does not allow a large 
deviation of non-extensive distribution from the usual MB distribution.
With a newly developed BBN code, the impact of primordial light-element abundances on $q$ values has been studied by utilizing the most recent BBN
cosmological parameters and the available nuclear cross-section data. For the first time, we have accurately verified the microscopic MB distribution 
with the macroscopic BBN theory and observation. By comparing the recent observed primordial abundances with our predictions,
only a tiny deviation of $\pm$6$\times$10$^{-4}$ at most can be allowed for the MB distribution. However, validity of the classical statistics needs to be 
studied further for the self-gravitating stars and binaries of high-density environment, with the extreme sensitivity of reverse rate on $q$ found here.
\end{abstract}



\maketitle
The Big-Bang theory is one of the pillars in the cosmological models, and the Big-Bang nucleosyntheis (BBN) model can test the fundamental physics as one
of the most powerful tools. Currently, the huge number of data coming from the astronomical observations and nuclear physics terrestrial experiments
represent that we have entered an era called precision cosmology. The key parameter, baryon-to-photon ratio $\eta$, in the standard BBN model has been
accurately constrained based on the observed anisotropy of the Cosmic Microwave Background (CMB) radiation from the Wilkinson Microwave Anisotropy Probe
(WMAP) satellite. The latest WMAP 9-years data~\cite{WMAP9} suggest $\Omega_Bh^{2}$=0.02264$\pm$0.0005, indicating an accurate value of
$\eta$=(6.203$\pm$0.137)$\times$10$^{-10}$.
The primordial abundances of the relic nuclei depend also on the early-Universe expansion rate~\cite{Steigman}. Any lepton asymmetry among neutral lepton 
(the neutrinos) can increase the neutrino energy densities and the expansion rate. In the standard model of particle physics, 
the number of light neutrino families, $N_{\nu}$=2.9840$\pm$0.0082, was determined by the recent CERN LEP experiments~\cite{Nv06}. Therefore, $N_{\nu}=3$ 
has been adopted in the current standard cosmology. 
In addition, the observational data of BBN relics have also been updated in recent years~\cite{bania02,Li2,aver13,cooke14}.

In BBN model, the major nuclear-physics inputs are the thermonuclear rates (or cross sections) of those reactions involving in the Big-Bang epoch. In the past
decades, great efforts have been undertaken to determine these data with high accuracy. The pioneering compilation of Fowler et al. in 1967~\cite{fcz67} led 
to the subsequent compilations~\cite{Wagoner67,wagoner69,fcz75,cf88,Smiths93} and a comprehensive compilation from the NACRE 
Collaboration~\cite{Angulo99}. Later, the astrophysical $S$-factors and rate parameters for those reactions of particular importance 
in BBN were compiled by Descouvemont et al.~\cite{Des04} and Serpico et al.~\cite{Serpico04}. Most recently, a newer reaction rate compilation similar to 
NACRE, called NACRE II, has been updated~\cite{yixu}.

BBN begins in earnest when the universe is 3-minutes old and ends less than half an hour later when the nuclear reactions are quenched by the low
temperature and density conditions in the expanding universe. Only those lightest nuclides (D, $^3$He, $^4$He, and $^7$Li) were synthesized in
the amounts of astrophysical interest. These relics provide us an unique window on the early universe. By comparing the predicted primordial abundances
with those inferred from the observational data, one can verify our standard models of cosmology and particles physics and, perhaps, discover clues to
modifications or extensions beyond them. Currently, the $\eta$-based BBN simulations give acceptable agreement between theoretical and observed
abundances of D and $^4$He, but it is still difficult to reconcile the predicted $^7$Li abundance with the observation. The BBN model overestimates
the primordial $^7$Li abundance by about a factor of three~\cite{Li2,Li}. Such discrepancy has promoted a wealth of experimental inquiries, and the accuracy 
of nuclear data used in the BBN reaction network has been greatly improved in recent years~\cite{Cyburt08}.
The conclusion is no clear nuclear physics resources capable of resolving the $^7$Li $puzzle$ within the standard BBN model.

In the current BBN network, thermonuclear reaction rates depend on the weighted thermal averages of the cross sections. The ion velocity distribution in 
the rate calculation is always assumed to be the classical Maxwell-Boltzmann (MB) 
distribution~\cite{mandl08,Rolfs88}. This distribution was first derived by James Clerk Maxwell~\cite{maxwell60} in 1860, and later Ludwig 
Boltzmann~\cite{boltz10} carried out significant investigations into the physical origins of this distribution in 1910.
It is well-known that MB distribution was derived for describing the thermodynamic equilibrium properties of the ideal gas (with mass density about order of 
10$^{-3}$ g/cm$^3$), where the particles move freely without interacting with one another, except for very brief collisions in which they exchange energy 
and momentum with each other or with their thermal environment. This classical distribution was ultimately verified with a high-resolution experiment by 
Miller and Kusch~\cite{miller55} at temperatures around 900 K in 1955. However, in real gases, there are various effects (e.g., van der Waals interactions, 
relativistic speed limits, etc.) that make their speed distribution sometimes very different from the Maxwell-Boltzmann form. 
Therefore, does the MB distribution still hold for the extremely high-temperature (in order of 10$^9$ K) plasma involved in the Big-Bang? 

The particle distribution in the space plasma is not purely Maxwellian, and Voronchev et al.~\cite{Nontherm08} supported the distribution of a fat
tail. In addition, the collisions between the energetic reaction products and thermal particles in the plasma, governed mainly by the nuclear force,
can transfer for a single event a large amount of energy and produce fast knock-on particles. Such collisions can enhance the high-energy tail of the
respective distributions, and such feature indeed exists in the hot plasma as demonstrated by Nakamura et al.~\cite{Nakamura06}.
Clayton et al.~\cite{Clyton75} once attempted to use the ion distribution in the solar with a depleted Maxwellian tail to solve
the famous solar neutrino problem, although which was later solved by the neutrino oscillations.
Later, Degl'Innocenti et al.~\cite{heli98} derived strong constraints on such deviations by using the detailed helioseismic information of the 
solar structure, and found the small deviation lying between -0.5\%$\thicksim$0.2\%. Recently, Bertulani et al.~\cite{Bertu13} has shown that only small 
deviation (-12\%$\thicksim$5\%) from the Maxwellian distribution is allowed for the BBN based on the Tsallis statistics introduced below. 



C. Tsallis introduced the concept of generalized non-extensive entropy and set up a new framework of statistics, called non-extensive statistics or Tsallis 
statistics~\cite{Tsallis88,Gell04,Tsal09}. It includes the Boltzmann-Gibbs (BG) statistics as a special limit of a more general theory. Nowadays, Tsallis 
statistics has been applied in a quite wide area, including physics, astronomy, biology, economics, etc. 
Especially, it has made the great success in the following aspects:
(i)the negative nature of heat capacity of thermaodynamically isolated self-gravitating systems depends on the non-extensive parameter $q$~\cite{Silva}; 
(ii)the velocity distribution of space plasmas can be well described by the non-Maxwellian distribution~\cite{Leubner04,velodis98}; 
(iii)the cosmic microwave background radiation of the universe can be fit better by non-extensive distribution than Gaussian distribution~\cite{Tsal09};
(iv)in particular, the non-extensive statistical effects are extremely relevant at the large baryon and energy densities as shown in the high-energy heavy-ion collisions~\cite{HIC}.

The nuclear reaction events occurring in the astrophysical environments are usually assumed to be independent and non-correlated.
The equilibrium properties of such a system are described by the equation-of-state of an ideal gas.
In fact, the classical MB distribution holds when the range of microscopic interactions
and memory are short. However, the stellar systems are generally subject to spatial long-range interactions, making thermodynamics of many-body
gravitating systems showing some peculiar features, such as negative specific heat and the absence of global entropy maxima, which are greatly different
from the usual thermodynamic systems. It has been thought that the BG statistics may be inappropriate to describe such systems.
In order to deal with such a problem, Tsallis proposed the generalization of the BG entropy, and introduced a parameter $q$, quantifying the 
degree of non-extensivity of the system (commonly referred to as the entropic index). $q$$>$1 leads to an entropy decreases, providing a state of higher order, 
whereas for $q$$<$1 the entropy increases, and the system evolves into disorder. Here, a $q$-Guassian ion-velocity distribution replaced the classical MB 
distribution in the frame work of Tsallis statistics.

In this Letter, we have investigated the impact of Tsallis $q$-Guassian distribution on the primordial BBN light-elements abundances. A series of new reaction 
(forward and reverse) rates for those important reactions of BBN interest have been calculated numerically by using $q$-Guassian and usual MB distributions 
without any approximations. 
We have implemented the reverse rate calculations with the $q$-Guassian distribution, and found for the first time that the reverse rates are extremely 
sensitive to the parameter $q$. Such feature does not allow the Tsallis distribution to deviate very much from the classical MB distribution.
By comparing the predicted BBN abundances with the recent astronomical observational data, we are aiming to verify the classical Maxwell-Boltzmann distribution, 
inaccessible to terrestrial experiment, by the BBN theory.


In Tsallis statistics, the $q$-Guassian three-dimensional velocity distribution is given by the formula~\cite{Silva2,Leubner04}
\begin{equation}
\label{eq:y1}
f_q(\mathbf{v})=B_q\left(\frac{m}{2\pi kT}\right)^{3/2}\left[1-(q-1)\frac{m\mathbf{v}^2}{2kT}\right]^{\frac{1}{q-1}}
\end{equation}
where $k$ is the Boltzmann constant. Here, $B_q$ denotes the $q$-dependent normalization constant given by
\begin{equation}
\label{eq:y2}
B_q=(q-1)^{1/2}\times\frac{3q-1}{2}\times\frac{q+1}{2}\times\frac{\Gamma(\frac{1}{2}+\frac{1}{q-1})}{\Gamma(\frac{1}{q-1})}
\end{equation} for $q$$>$1, and
\begin{equation}
\label{eq:y3}
B_q=(1-q)^{3/2}\times\frac{\Gamma(\frac{1}{1-q})}{\Gamma(\frac{1}{1-q}-\frac{3}{2})}
\end{equation} for $q\leqslant1$.
Note that there exists a thermal cutoff on the maximum velocity allowed for a particle in the $q$$>$1 case, i.e., $v_\mathrm{max}=\sqrt{2kT/m(q-1)}$. 
In the limit of $q\rightarrow1$, $f_q$ reduces to the standard MB distribution
\begin{equation}
\label{eq:y4}
f(\mathbf{v})=\left(\frac{m}{2\pi kT}\right)^{3/2}\mathrm{exp}\left(-\frac{m\mathbf{v}^2}{2kT}\right).
\end{equation}
Utilizing $E=\frac{1}{2}mv^2$, we can convert the non-extensive velocity distribution function $f_q(v)$ into energy distribution function $f_q(E)$,
\begin{widetext}
\begin{eqnarray}
\label{eq:y5}
f_q(E)=2B_q\left[\frac{E}{\pi(kT)^3}\right]^{1/2}\left[1-(q-1)\frac{E}{kT}\right]^{\frac{1}{q-1}} \overset{q\rightarrow1}{\longrightarrow}2\left[\frac{E}{\pi(kT)^3}\right]^{1/2}\mathrm{exp}\left(-\frac{E}{kT}\right).
\end{eqnarray}
\end{widetext}

It is well-known that thermonuclear rate for a typical $1+2\rightarrow3+4$ reaction can be calculated by folding the cross section $\sigma(E)$ with a MB 
distribution~\cite{Rolfs88}
\begin{equation}
\label{eq:z}
\left\langle\sigma v\right\rangle_{12}=\sqrt{\frac{8}{\pi\mu_{12}(kT)^3}}\int_{0}^{\infty}\sigma(E)_{12}E\mathrm{exp}\left(-\frac{E}{kT}\right)\,dE,
\end{equation}
with $\mu_{12}$ the reduced mass of particles $1$ and $2$, and $\sigma(E)_{12}$ the corresponding cross section. With the non-extensive distribution, 
the reaction rate becomes
\begin{widetext}
\begin{equation}
\label{eq:2}
\left\langle\sigma v\right\rangle_{12}=B_q\sqrt{\frac{8}{\pi\mu_{12}}}\times\frac{1}{(kT)^3}\int_{0}^{E_\mathrm{max}}\sigma_{12}(E)E\left[1-(q-1)\frac{E}{kT}\right]^{\frac{1}{q-1}}\,dE.
\end{equation}
\end{widetext}
with $E_\mathrm{max}$=$\frac{kT}{q-1}$ for $q$$>$1, and +$\infty$ for 0$<$$q$$<$1. Here, we exclude the situation of $q$$<$0 according to the 
maximum-entropy principle~\cite{Tsallis88,Tsal09}.

In MB distribution frame, the ratio between reverse rate and forward one is found to be 
$\left\langle\sigma v\right\rangle_{34}$/$\left\langle\sigma v\right\rangle_{12}$=$c\times\mathrm{exp}(-\frac{Q}{kT})$,
with a constant factor defined as $c=\frac{(2J_1+1)(2J_2+1)(1+\delta_{34})}{(2J_3+1)(2J_4+1)(1+\delta_{12})}\left(\frac{\mu_{12}}{\mu_{34}}\right)^{3/2}$.
In fact, the Tsallis distribution (rather than the MB distribution) should be coherently applied to the reverse rate calculation. Here, we have calculated 
the reverse rate numerically with the Tsallis distribution by the following equation:
\begin{widetext}
\begin{equation}
\label{eq:rev}
\left\langle\sigma v\right\rangle_{34}=c\times B_q\sqrt{\frac{8}{\pi\mu_{12}}}\times\frac{1}{(kT)^3}\int_{0}^{E_\mathrm{max}-Q}\sigma_{12}(E)E\left[1-(q-1)\frac{E+Q}{kT}\right]^{\frac{1}{q-1}}\,dE.
\end{equation}
\end{widetext}

\begin{figure}[tbp]
\begin{center}
\includegraphics[width=8.5cm]{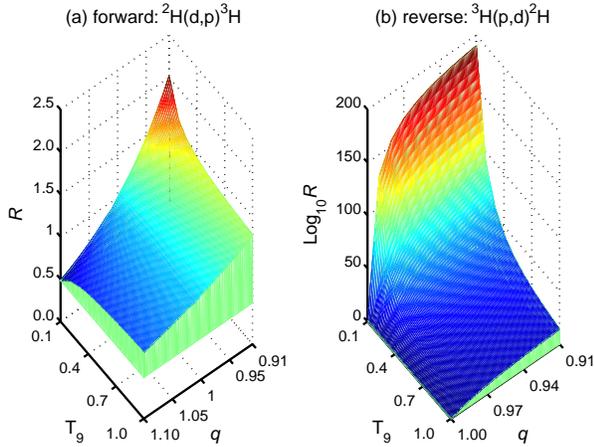}
\vspace{-3mm}
\caption{\label{fig1} Ratio between the Taslli and MB distributions for the $^2$H(d,p)$^3$H reaction as functions of temperature $T_9$ and $q$ values,
(a) for forward reaction (in linear scale), and (b) for reverse reaction (in logarithm scale).}
\end{center}
\end{figure}

\begin{figure}[tbp]
\begin{center}
\includegraphics[width=8.5cm]{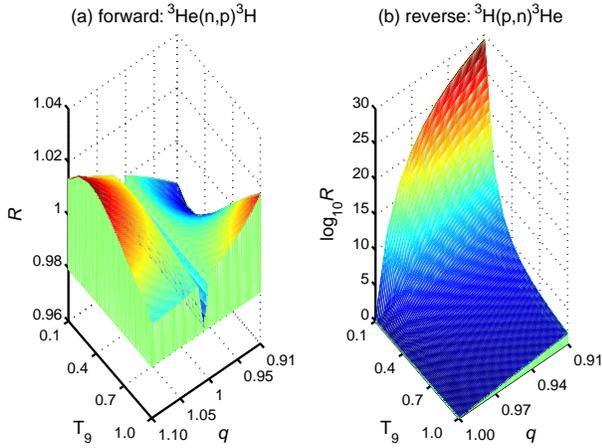}
\vspace{-3mm}
\caption{\label{fig2} Results for the $^3$He(n,p)$^3$H reaction, see caption of Fig.~\ref{fig1}.}
\end{center}
\end{figure}

\begin{figure}[tbp]
\begin{center}
\includegraphics[width=7.0cm]{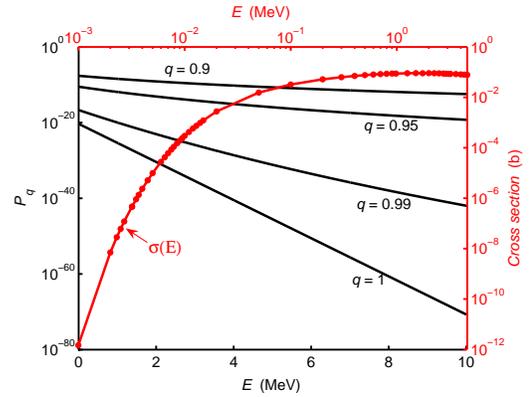}
\vspace{-3mm}
\caption{\label{fig3} Illustration of $q$ sensitivity to reverse rate of the $^{2}$H(d,p)$^{3}$H reaction.}
\end{center}
\end{figure}

We show below the impact of $q$ values on the forward and reserve rates of two types of reactions. Here, $^{2}$H(d,p)$^{3}$H is taken as an example of 
the charged-particle-induced reaction, and $^3$He(n,p)$^3$H as that of the neutron-induced reaction. Both are among the most important reactions involved 
in the BBN. The ratios ($R$) between reaction rates operated with the Tsallis-distribution and MB-distribution are calculated for these two reactions.
Figs.~\ref{fig1} and~\ref{fig2} show the results of forward and reverse rates as functions of temperature and $q$ value, i.e., in unit of the usual MB rates. 
Here, the cross section data for these two reaction are compiled in Refs.~\cite{Angulo99,Des04}. 
In region of 0.1$\leq$$T_9$$\leq$1.0 and 0.91$\leq$$q$$\leq$1.1, the forward rates calculated with the Tsallis-distribution deviate from the MB rates by 
about factors of 2 and 2\% (at maximum) for the $^{2}$H(d,p)$^{3}$H and $^3$He(n,p)$^3$H reactions, respectively. It means that the changes in the forward rates 
are ``moderate" within a small deviation of $\pm$10\%. However, the reverse rates of both types of reaction are extremely sensitive to $q$ value. 
For 0.91$\leq$$q$$\leq$1 (i.e., $q$$<$1), the corresponding reverse rates (with Tsallis-distribution) deviate
tremendously from those MB rates by about 200 and 30 orders of magnitude for $^{2}$H(d,p)$^{3}$H and $^3$He(n,p)$^3$H reactions, respectively. 
For instance, even with a very small deviation ($q$=0.999), the reverse rates of $^{2}$H(d,p)$^{3}$H deviates from the MB ones by about factors of 
4$\times$10$^{10}$ at 0.2 GK, and 3.3 at 1.0 GK. 
Here, the reverse rates with $q$$>$1 are not shown because they are negligible in comparison with the MB rates.
Fig.~\ref{fig3} shows schematically the $q$ sensitivity to reverse rate of the $^{2}$H(d,p)$^{3}$H reaction. 
The probability distribution shown is defined as $P_q(E)$=[1-($q$-1)$\frac{E+Q}{kT}$]$^{\frac{1}{q-1}}$, which contains in the integral of Eq.~\ref{eq:rev}. 
The sensitivity of 
$P_q(E)$ (i.e., the tail of distribution) on $q$ results in the huge deviations for the reverse rates comparing to those MB rates.
The extreme sensitivity of reverse rate on parameter $q$ has an very important impact: for $q$$<$1, the reverse rates are greatly increased comparing to the 
forward ones, meaning that the nucleus can be destructed very effectively and BBN cannot go further; while for $q$$>$1, the reverse rates become negligible 
comparing to the forward ones, an opposite effect to $q$$<$1.

Recently, a new BBN code~\cite{bbn} has been developed on the base of refs.~\cite{Wagoner67,wagoner69}.
As we know that there are totally 11 reactions~\cite{Smiths93} are of primary importance in the primordial light-element production. 
Therefore, the present reaction network only involves those nuclides of $A\leqslant9$ (see Table~\ref{tab1}), quite different from the previous big 
one~\cite{coc12}. In addition, the sub-leading reactions indicated in ref.~\cite{Serpico04} are also included. Thus, our network is sufficient to study the 
primordial abundances of D, $^3$He, $^4$He and $^7$Li.
In order to save the computational cost, we have only performed the non-extensive operation for those important reactions (with a bold face in
Table~\ref{tab1}), and the available compiled MB rates are adopted for the remaining reactions. In addition, the updated neutron lifetime of 
$\tau_n=887.7$ s~\cite{nlife} is utilized in our model calculation.
Our predicted primordial BBN abundances with the usual MB distribution (i.e. $q$=1) are listed in Table~\ref{tab2}. The predictions by Bertulani
et al.~\cite{Bertu13} (with MB distribution) and Coc. et al.~\cite{coc12}, as well as the recent observational data are listed for comparison.
Our results are well consistent with those previous predictions, and also agree well with the observations, except for the $^7$Li abundance.
This consistency ensures the correctness of the present BBN calculations.

\begin{table}[t]
\centering
\setlength{\belowcaptionskip}{10pt}
\caption{\label{tab1} Nuclear reactions involved in the present BBN network. The non-extensive Tsallis distribution is implemented for 17 reactions shown in 
the bold face (11 of primary importance in italic and 6 of secondary importance). The references for the reaction data adopted are also listed.}
\begin{tabular}{|ll|ll|}
\hline
Reaction & Ref. & Reaction & Ref. \\
\hline
(1) n $\to$ p     & \cite{nlife}                                 & (18) \textbf{$^2$H($\alpha,\gamma$)$^6$Li} & \cite{Angulo99,yixu} \\
(2) $^3$H$\to^3$He & \cite{Tdecay}                               & (19) \textbf{\textit{$^3$H($\alpha,\gamma$)$^7$Li}} & \cite{Des04}\\
(3) $^8$Li$\to$2$^4$He & \cite{Tilley04}                           & (20) \textbf{\textit{$^3$He($\alpha,\gamma$)$^7$Be}} & \cite{Des04}\\
(4) $^6$He$\to^6$Li  & \cite{Tilley02}                              & (21) \textbf{\textit{$^2$H(d,n)$^3$He}} & \cite{Des04}\\
(5) \textbf{\textit{$^1$H(n,$\gamma$)$^2$H}}  & \cite{npD}                          & (22) \textbf{\textit{$^2$H(d,p)$^3$H}} & \cite{Des04}\\
(6) $^3$He(n,$\gamma$)$^4$He  & \cite{wagoner69}                     & (23) \textbf{\textit{$^3$H(d,n)$^4$He}} & \cite{Des04}\\
(7) $^6$Li(n,$\gamma$)$^7$Li & \cite{MFowler}                     & (24) \textbf{\textit{$^3$He(d,p)$^4$He}} & \cite{Des04}\\
(8) \textbf{\textit{$^3$He(n,p)$^3$H}}    & \cite{Des04}                          & (25) $^3$He($^3$He,2p)$^4$He & \cite{cf88} \\
(9) \textbf{\textit{$^7$Be(n,p)$^7$Li}}    & \cite{Des04}                         & (26) \textbf{$^7$Li(d,n)2$^4$He} & \cite{cf88}\\
(10) $^2$H(n,$\gamma$)$^3$H   & \cite{wagoner69}                    & (27) \textbf{$^7$Be(d,p)2$^4$He} & \cite{Paker72,cf88}\\
(11) $^6$Li(n,$\alpha$)$^3$H  & \cite{cf88}                    & (28) $^7$Li(n,$\gamma$)$^8$Li & \cite{wagoner69}\\
(12) \textbf{$^7$Be(n,$\alpha$)$^4$He} & \cite{King77}                    & (29) $^9$Be(p,$\alpha$)$^6$Li & \cite{cf88}\\
(13) \textbf{\textit{$^2$H(p,$\gamma$)$^3$He}}  & \cite{Des04}                    & (30) 2$^4$He(n,$\gamma$)$^9$Be & \cite{cf88}\\
(14) \textbf{$^3$H(p,$\gamma$)$^4$He}  & \cite{PTHe4}                    & (31) $^8$Li(p,n)2$^4$He & \cite{wagoner69}\\
(15) $^6$Li(p,$\gamma$)$^7$Be & \cite{yixu}                    & (32) $^9$Be(p,d)2$^4$He & \cite{cf88}\\
(16) \textbf{$^6$Li(p,$\alpha$)$^3$He} & \cite{Angulo99,yixu}                    & (33) $^8$Li(n,$\gamma$)$^9$Li & \cite{ZHLi}\\
(17) \textbf{\textit{$^7$Li(p,$\alpha$)$^4$He}} & \cite{Des04}                    & (34) $^9$Li(p,$\alpha$)$^6$He & \cite{Thomas93}\\
\hline
\end{tabular}
\end{table}

\begin{table}[tbp]
\centering
\setlength{\belowcaptionskip}{10pt}
\caption{\label{tab2} The predicted abundances for the BBN primordial light elements with the usual MB distribution (i.e. $q$=1). The observational data 
are listed for comparison.}
\begin{tabular}{|c|c|c|c|c|}
\hline
Abundance & Present & Ref.~\cite{Bertu13} & Ref.~\cite{coc12} & Observation \\
\hline
$^4$He                       & 0.2485 & 0.249 & 0.2476 & 0.2464$\pm$0.0097~\cite{aver13} \\
D/H($\times$10$^{-5}$)       & 2.54   & 2.62  & 2.59   & 2.53$\pm$0.04~\cite{cooke14} \\
$^3$He/H($\times$10$^{-5}$)  & 1.01   & 0.98  & 1.04   & 1.1$\pm$0.2~\cite{bania02} \\
$^7$Li/H($\times$10$^{-10}$) & 5.34   & 4.39\footnote{This value deviates considerably with that in Ref.~\cite{coc12} and ours.}  & 5.24   & 1.58$\pm$0.31~\cite{Li2} \\
\hline
\end{tabular}
\end{table}

\begin{figure}[tbp]
\begin{center}
\includegraphics[width=7.0cm]{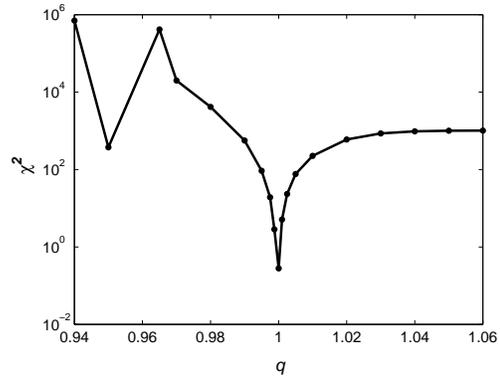}
\vspace{-3mm}
\caption{\label{fig4} Calculated $\chi^2$ as a function of non-extensive parameter $q$ by using the observed abundance data of D/H, $^3$He/H, and $^4$He.}
\end{center}
\end{figure}

The chi-square ($\chi^2$) fit has been performed to search for an appropriate $q$ value with which one can well reproduce the observed primordial abundances. 
Here, the abundances data observed for the D/H, $^3$He/H and $^4$He nuclides listed in Table~\ref{tab2} are adopted, while the 
$^7$Li data are excluded since the current BBN model yet solves the lithium \emph{puzzle}. The $\chi^2$ is defined by the minimization of
\begin{equation}
\label{eq:fit}
\chi^2=\sum_{i}\left[\frac{Y_i(q)-Y_i(obs)}{\sigma_i}\right]^2,
\end{equation}
where $Y_i$($q$) is the abundance (of nuclide $i$) predicted with a non-extensive parameter $q$, and $Y_i(obs)$ is the observed ones with $\sigma_i$ the 
observational error. $\chi^2$ is plotted in Fig.~\ref{fig4} as a function of parameter $q$ varying from 0.94 to 1.06 (i.e., deviation of $\pm$6\%). 
For the extreme case of $q$=0.5 and 2.0 used in the work of Bertulani et al., the BBN results are meaningless. The range of $q$ selected here is owing to 
the sensitive of the reverse rate discussed above. The sensitive found here has a very interesting consequence: the Tsallis distribution cannot be
allowed to deviate very much from the classical MB distribution for the Big-Bang plasma.
It shows gracefully that the $\chi^2$ function is minimized at unity (i.e., $q$=1). The D/H, $^3$He/H and $^4$He abundances fall into the observational 
regions (1$\sigma$ error), at most with deviations of (-0.06$\thicksim$0.01)\%, (-0.8$\thicksim$0.5)\% and $\pm$0.2\%, respectively. With the strongest 
observational constraint on deuteron abundance, it can be concluded that the deviation from the MB distribution can be at most $\pm$0.06\% for the Big-Bang 
plasma. Comparing to the possible deviation (-12\%$\thicksim$5\%) estimated by Bertulani et al.~\cite{Bertu13}, our constraint is much more stringent.
Thus, we have accurately verified the classical MB distribution for the extreme environment of Big-Bang plasma, and a level of $\pm$6$\times$10$^{-4}$ 
at most can be allowed for the deviation.

The present results are different from those of Bertulani et al. in three aspects:
(i) the forward rates are different. They operated the integral in a narrow energy region ($\pm$5$\Delta E_0$) which is not sufficient for a 
big $q$ value (e.g., 0.5 and 2.0), while we have integrated the cross section up to the maximum energy experimentally achieved;
(ii) the constraints on $q$ are different, mainly because they did not operate the reverse rate calculations with the non-extensive 
distribution coherently;
(iii) the evolving times ($t_\mathrm{peak}$) for the D/H, $^3$He/H and $^4$He abundances reaching the maximum are considerably different. 
The presently calculated primordial abundances (with $q$=1) with respect to the evolving time agree very well with those of Coc et al.~\cite{coc12}. 
However, for example, the $t_\mathrm{peak}$ for D/H calculated by Bertulani et al. is earlier about 80 s comparing with that of Coc et al. and ours. 
There might be some problems in the previous work of Bertulani et al. The detailed comparison and explanation will be published elsewhere~\cite{housq}.


In fact, massive objects usually occur Gravothermal catastrophe, i.e. thermodynamic instability due to negative specific heat~\cite{Lynden68}, forming a
core-halo structure, well known from the observations. Such a stellar plasma system with a velocity space structure generally characterized by the different
densities and temperatures of core and halo populations, approaching a superposition of two equilibrium distributions~\cite{Leubner04}. A typical
feature is that the leptokurtic and long-tailed probability distribution functions subject to a non-Gaussian core and a pronounced suprathermal halo.
Here, the core corresponds to the $q$$>$1 case, and halo to the $q$$<$1 case.
However, the results from the 2dF Galaxy Redshift survey~\cite{Nature} and recent CMB temperature from the WMAP-7 years data~\cite{wmap7}, which show the matter 
distribution in the universe is uniform and isotropic, rule out the probability of core-halo structure in the Big-Bang.  
Therefore, unlike self-gravitating stars the primordial plasma fits well the MB distribution, notwithstanding both of them are belonging to
the self-gravitating systems. 
The reason is summarized as below:
BBN occurs in a high entropy and low density (baryon density about 10$^{-5}$--10$^{-8}$ g/cm$^3$~\cite{wagoner69}) environment, where particles are energetic 
and interact very frequently. The driving power for the expansion is extremely strong relative to the gravitation in spite of knowing nothing about the origin 
of this power. Facing so rapid expansion, the gravitation can not induce particle cluster, and no violent change takes place in gravity field as occurs in stars. 
In BBN environment, the relax time is so short that the system is extremely easy to achieve thermal equilibrium, and the baryon distribution is also perfectly 
homogenous (no core) relative to that in stars. However, validity of the classical statistics still needs to be studied further for the self-gravitating stars 
(e.g., density about 150 g/cm$^3$~\cite{solar} in the core of our Sun), as well as those close binary systems 
(with high density about 10$^{3}$--10$^{8}$ g/cm$^3$~\cite{cham92}), e.g., nova, x-ray bursts and Type I supernova.

\begin{acknowledgments}
This work was financially supported by the Major State Basic Research Development Program of China (2013CB834406) and the National Natural Science Foundation of 
China (Nos. 11135005, 11321064).
\end{acknowledgments}

\end{document}